\documentclass[letterpaper]{article} % DO NOT CHANGE THIS
\usepackage{aaai24}  % DO NOT CHANGE THIS
\usepackage{times}  % DO NOT CHANGE THIS
\usepackage{helvet}  % DO NOT CHANGE THIS
\usepackage{courier}  % DO NOT CHANGE THIS
\usepackage[hyphens]{url}  % DO NOT CHANGE THIS
\usepackage{graphicx} % DO NOT CHANGE THIS
\usepackage{natbib}  % DO NOT CHANGE THIS AND DO NOT ADD ANY OPTIONS TO IT
\usepackage{caption} % DO NOT CHANGE THIS AND DO NOT ADD ANY OPTIONS TO IT
\usepackage{algorithm}
\usepackage{algorithmic}

\usepackage[hypertexnames=false]{hyperref}
\usepackage{mathtools}
\usepackage{color}
\usepackage{amsthm}
\usepackage{amsfonts}
\newcommand{\bs}{\boldsymbol}
\newcommand{\eq}{{\rm st}}
\newcommand{\norm}{{\rm Norm}}

\newcommand{\mc}{\mathcal}
\newtheorem{theorem}{Theorem}
\newtheorem{assumption}{Assumption}
\newtheorem{definition}{Definition}
\newtheorem{corollary}{Corollary}

\usepackage{comment}

\newcommand{\rd}{\mathrm{d}}
\newcommand{\eqcon}{*}
\usepackage{newfloat}
\usepackage{listings}
\DeclareCaptionStyle{ruled}{labelfont=normalfont,labelsep=colon,strut=off} % DO NOT CHANGE THIS
\floatstyle{ruled}
\newfloat{listing}{tb}{lst}{}
\floatname{listing}{Listing}
\title{Memory Asymmetry Creates Heteroclinic Orbits to Nash Equilibrium\\ in Learning in Zero-Sum Games\footnote{The codes that we used are available at \protect\url{https://github.com/CyberAgentAILab/learning_games_with-memory_asymmetry}}}
\author{
    Yuma Fujimoto\textsuperscript{\rm 1,2,3}, Kaito Ariu\textsuperscript{\rm 3,4}, Kenshi Abe\textsuperscript{\rm 3,5}\\
}
\affiliations{
    \textsuperscript{\rm 1}SOKENDAI, \textsuperscript{\rm 2}The University of Tokyo, \textsuperscript{\rm 3}CyberAgent, \textsuperscript{\rm 4}KTH, \textsuperscript{\rm 5}The University of Electro-Communications\\
    \protect\url{fujimoto_yuma_xa@cyberagent.co.jp}
    
}

\usepackage{bibentry}
\begin{document}

\maketitle

\begin{abstract}
Learning in games considers how multiple agents maximize their own rewards through repeated games. Memory, an ability that an agent changes his/her action depending on the history of actions in previous games, is often introduced into learning to explore more clever strategies and discuss the decision-making of real agents like humans. However, such games with memory are hard to analyze because they exhibit complex phenomena like chaotic dynamics or divergence from Nash equilibrium. In particular, how asymmetry in memory capacities between agents affects learning in games is still unclear. In response, this study formulates a gradient ascent algorithm in games with asymmetry memory capacities. To obtain theoretical insights into learning dynamics, we first consider a simple case of zero-sum games. We observe complex behavior, where learning dynamics draw a heteroclinic connection from unstable fixed points to stable ones. Despite this complexity, we analyze learning dynamics and prove local convergence to these stable fixed points, i.e., the Nash equilibria. We identify the mechanism driving this convergence: an agent with a longer memory learns to exploit the other, which in turn endows the other's utility function with strict concavity. We further numerically observe such convergence in various initial strategies, action numbers, and memory lengths. This study reveals a novel phenomenon due to memory asymmetry, providing fundamental strides in learning in games and new insights into computing equilibria.
\end{abstract}

\section{Introduction}
Learning in games discusses how players learn their strategies through repeated games~\cite{fudenberg1998theory}. Here, whether or not players learn their optimal strategies, called Nash equilibrium, is a nontrivial problem because each player's optimal strategy depends on the strategy of his/her opponent. This problem becomes further exacerbated in zero-sum games, where both the players conflict in their utility functions. In order to approach this problem, various learning algorithms have been proposed, such as replicator dynamics~\cite{borgers1997learning, hofbauer1998evolutionary, sato2002chaos} and gradient ascent~\cite{singh2000nash, zinkevich2003online, bowling2002multiagent, bowling2004convergence}. The dynamics of these algorithms have undergone extensive study, with findings indicating a cycling behavior around the Nash equilibrium in zero-sum games~\cite{mertikopoulos2016learning, mertikopoulos2018cycles}.

Memory, referring to an agent's ability to alter their subsequent actions based on past decisions, is often incorporated into games to investigate more sophisticated strategies and explore the decision-making processes of real agents, such as humans. Such memory introduces greater complexity and diversity into games.
A celebrated study in economics has shown that players with memory can take various strategies in the Nash equilibria~\cite{fudenberg2009folk}. Indeed, in prisoner's dilemma games, agents with memory can achieve cooperation~\cite{axelrod1981evolution} and asymmetric degree of cooperation~\cite{fujimoto2019emergence}, even though those without memories cannot cooperate in the Nash equilibrium. Memory also complicates learning dynamics such as chaotic dynamics~\cite{barfuss2019deterministic} and divergence from the Nash equilibrium~\cite{fujimoto2023learning} in zero-sum games. 
In recent years, theoretical approaches to games with memories have garnered significant interest~\cite{barfuss2020reinforcement, usui2021symmetric, meylahn2022limiting, ueda2023memory}.

How learning changes depending on memory length is an interesting question in games with memory. Indeed, this question has often been discussed in prisoner's dilemma games. It is reported that agents with longer memories can cooperate more cleverly~\cite{hilbe2017memory, murase2020five}. Furthermore, the difference in payoffs emerges between agents with different memory lengths in various learning algorithms such as Q-learning~\cite{sandholm1996multiagent}, coupled replicator dynamics~\cite{fujimoto2021exploitation}, and evolutionary dynamics~\cite{baek2016comparing, schmid2022direct}. This suggests that an asymmetry in memory capacities could potentially lead to the exploitation of the opponent's payoff in zero-sum games. Despite this, the impact of memory asymmetry in games has yet to be thoroughly investigated.

This study considers learning in zero-sum games between agents with memory asymmetry. First, we formulate games with asymmetry memory and a gradient ascent algorithm in the games. This algorithm corresponds to replicator dynamics, provided that the learning rate is sufficiently small, as outlined in~\citet{fujimoto2023learning}.
In order to catch theoretical insights, we first focus on two-action zero-sum games between agents with memory lengths of one and zero (called one-memory and zero-memory). We analyze the Nash equilibrium in this with-memory game in comparison with the ``original'' Nash equilibrium in games without memory. Interestingly, we prove that this original Nash equilibrium can be divided into two distinct regions. The first region comprises unstable fixed points within the learning dynamics, while the second consists of stable fixed points.
These stable points correspond to the Nash equilibrium in with-memory games. 
Thus, heteroclinic orbits~\cite{strogatz2018nonlinear} are observed in with-memory games; Dynamics diverge from the unstable region and converge to the stable region within the original Nash equilibrium. This convergence is because the agent with a longer memory learns to concave the opponent's utility function. Furthermore, we experimentally confirmed this convergence across various initial states, action numbers, and memory lengths.

\begin{figure*}[ht]
    \centering
    \includegraphics[width=0.7\hsize]{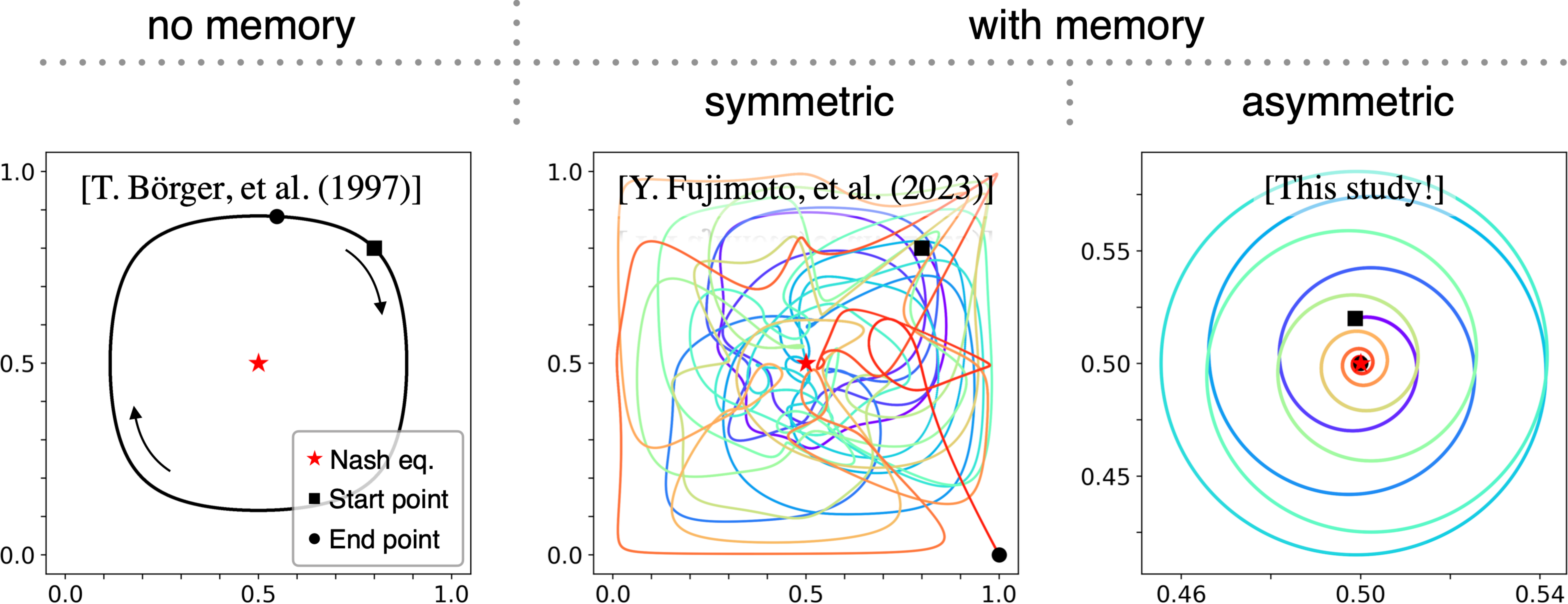}
    \caption{Learning dynamics illustrated for three memory-configuration scenarios involving two agents. Learning dynamics show a cycling behavior around the Nash equilibrium when the agents have no memory (left panel). Learning dynamics diverge from the Nash equilibrium when the agents have the same memory capacity (center). Learning dynamics draw heteroclinic orbits and eventually converge to the Nash equilibrium when the agents have different memory lengths (right). In all the panels, the horizontal and vertical axes indicate the probabilities of the agents choosing ``head'' in matching-pennies games (see Fig.~\ref{F02}). In the center and right panels, the color gradient indicates the passage of time (blue represents older data, and red represents newer data).}
    \label{F01}
\end{figure*}

To summarize, our study uncovers complex heteroclinic dynamics in learning in with-memory games. We also identify a novel phenomenon of convergence in learning in zero-sum games, which is achieved due to memory asymmetry. Given that replicator dynamics exhibit cyclical behavior without memory (see Panel A in Fig.~\ref{F01}) and diverging behavior with memory symmetry (see Panel B), our discovery is significant (see Panel C).
% Figure 01

\section{Preliminary}
\subsection{Two-Player Normal-Form Games}
\begin{figure*}[ht]
    \centering
    \includegraphics[width=0.8\hsize]{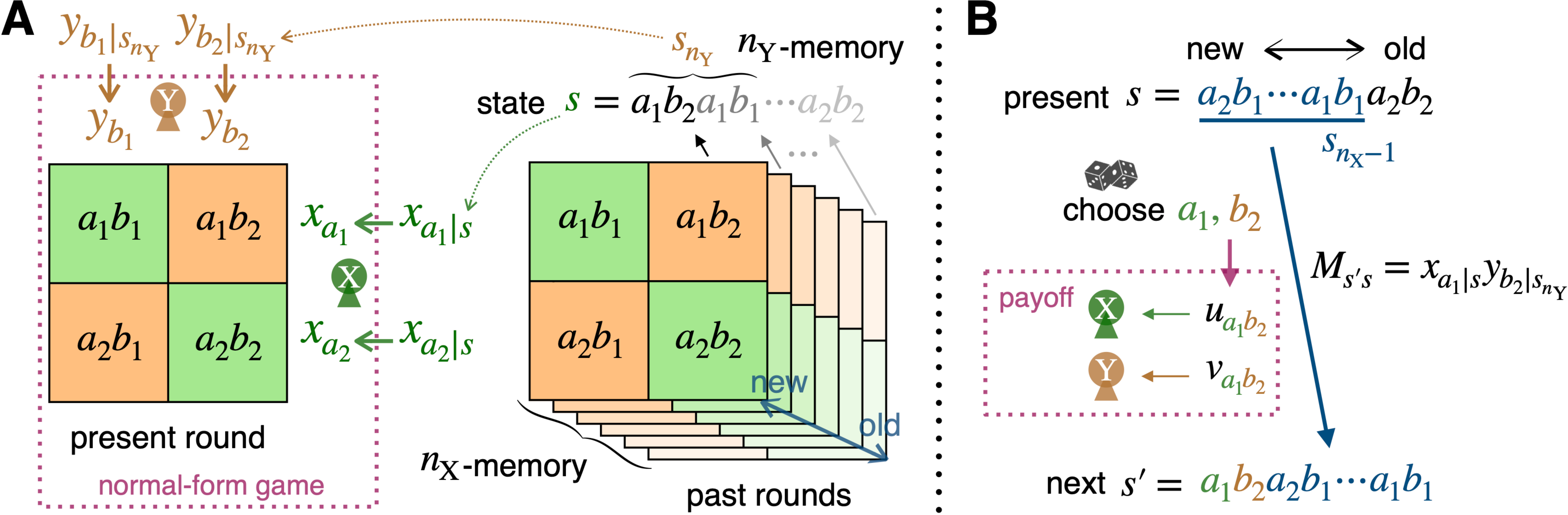}
    \caption{A: Schematics of with-memory games. The area surrounded by the magenta dotted line shows a classic normal-form game, where player X (green) chooses action either $a_1$ or $a_2$ with the probability of $x_{a_1}$ and $x_{a_2}$ in the row of the matrix, while player Y (orange) chooses action either $b_1$ or $b_2$ with the probability of $y_{b_1}$ and $y_{b_2}$ in the column. Especially in the matching-pennies game, $a_1=b_1=$ ``head'' and $a_2=b_2=$ ``tail''. The matching of X's and Y's actions leads to X's win (green panel), while the mismatching leads to Y's win (orange panel). In with-memory games, $x_{a_i}$ and $y_{b_i}$ is given by $x_{a_i|s}$ and $y_{b_i|s_{n_{\rm Y}}}$. Here, $s$ is the string of their actions played in the previous $n_{\rm X}$ rounds. In addition, because Y has a shorter memory than X, $s_{n_{\rm Y}}$ is defined as the substring of $s$. B: Schematics of Markov transition in with-memory games. In the transition from $s$ to $s'$, $s_{n_{\rm X}-1}$ (blue), i.e., the last $2(n_{\rm X}-1)$ substring of $s$ continues to exist in $s'$. X and Y choose actions $a_1$ (green) and $b_2$ (orange) are appended to this substring $s_{n_{\rm X}-1}$. These choices occur with the probability of $M_{s's}$ and give X and Y the payoffs of $u_{a_1b_2}$ and $v_{a_1b_2}$, respectively.}
    \label{F02}
\end{figure*}

We consider two players of X and Y. In every round, each of them chooses its action from $\mc{A}=\{a_1,\cdots,a_m\}$ and $\mc{B}=\{b_1,\cdots,b_m\}$. If they choose $a\in\mc{A}$ and $b\in\mc{B}$, they immediately receive a payoff of $u_{ab}\in\mathbb{R}$ and $v_{ab}\in\mathbb{R}$, respectively. For the illustration of two-player normal-form games, see the area surrounded by the magenta dotted lines in Fig.~\ref{F02}-A.
% Figure 02

\subsection{Games with Memory Asymmetry}
We assume that X and Y can memorize their actions in the newest $n_{\rm X}\in\mathbb{N}$ and $n_{\rm Y}\in\mathbb{N}$ rounds, respectively. 
We set that X has a longer memory ($n_{\rm X}>n_{\rm Y}$). 
Let $\mc{S}=\prod_{k=1}^{n_{\rm X}}(\mc{A}\times\mc{B})$ be the set of all states that X can memorize. 
Notably, each state $s\in\mc{S}$ is given by a string of actions of length $2n_{\rm X}$. Under state $s$, X can choose its action $a$ with the probability of $x_{a|s}\in(0,1)$. X's strategy is the set of $x_{a|s}$ for all $a\in\mc{A}$ and $s\in\mc{S}$ and is denoted as $|\mc{S}|(=m^{2n_{\rm X}})$-numbers of $(m-1)$-dimension simplexes, ${\bf x}\in\prod_{s\in\mc{S}}\Delta^{m-1}$. Fig.~\ref{F02}-A shows the introduction of memories into normal-form games.

For each $n\in\{0,\ldots,n_{\rm X}-1\}$, let $s_{n}$ denote the newest $n$ substring of $s$. We also define the set $\mc{S}_{n}$ as $\mc{S}_{n}=\prod_{k=1}^{n}(\mc{A}\times\mc{B})$, which represents the set of all possible $s_{n}$. Since the length of Y's memory is $n_{\rm Y}(<n_{\rm X})$, the set of states memorized by Y can be represented by $\mc{S}_{n_{\rm Y}}$. We assume that under each state $s_{n_{\rm Y}}$, Y chooses its action $b$ with probability $y_{b|s_{n_{\rm Y}}}\in(0,1)$. If $n_{\rm Y}=0$, $\mc{S}_{n_{\rm Y}}=\emptyset$ and we simply denote $y_{b|s_{n_{\rm Y}}}$ as $y_{b}$. Y's strategy is given by ${\bf y}\in\prod_{s_{n_{\rm Y}}\in\mc{S}_{n_{\rm Y}}}\Delta^{m-1}$. Because the above immediate payoff is determined by both the players' actions in the last round ($s_1\in\mc{A}\times\mc{B}$), we can rewrite the payoff as $u_{s_1}$ and $v_{s_1}$.

\subsection{Formulation as Markov Transition Processes}
The above setting is described by a discrete-time Markov transition process (see the illustration of Fig.~\ref{F02}-B), where the transition rate from state $s$ to $s'$ is
\begin{align}
    M_{s's}({\bf x}, {\bf y}):=\begin{cases}
        x_{a|s}y_{b|s_{n_{\rm Y}}} & (s'=abs_{n_{\rm X}-1}) \\ % s_{-}
        0 & ({\rm otherwise}) \\
    \end{cases}.
    \label{matrix}
\end{align}

We formulate the Nash equilibrium based on the Markov transition process. 
Then, strategies ${\bf x}$ and ${\bf y}$ may be regarded as fixed in the above Markov transition process. For fixed ${\bf x}$ and ${\bf y}$ within the interior of the simplexes, the stationary state for the above Markov transition processes is uniquely denoted as $p_{s}^{\eq}({\bf x}, {\bf y})$. The stationary state satisfies $p_{s'}^{\eq}({\bf x}, {\bf y})=\sum_{s}M_{s's}({\bf x}, {\bf y})p_{s}^{\eq}({\bf x}, {\bf y})$. If the game continues a sufficiently long time, each player's payoff is given under this stationary state; $u^{\eq}({\bf x}, {\bf y})=\sum_s p_{s}^{\eq}({\bf x}, {\bf y})u_{s_{1}}$ and $v^{\eq}({\bf x}, {\bf y})=\sum_s p_{s}^{\eq}({\bf x}, {\bf y})v_{s_{1}}$. Therefore, the above payoffs reflect that every player learns slowly enough, i.e., their learning rates are sufficiently small. The Nash equilibrium is defined as the strategies that maximize their own payoff functions in the stationary state;
\begin{align}
    \begin{cases}
        {\bf x}^*\in\mathrm{argmax}_{{\bf x}} u^{\eq}({\bf x}, {\bf y}^*)\\
        {\bf y}^*\in\mathrm{argmax}_{{\bf y}} v^{\eq}({\bf x}^*, {\bf y})\\
    \end{cases}.
\end{align}
The objective of this study is to learn this Nash equilibrium.

\section{Algorithm}
Referring to~\citet{fujimoto2023learning}, we now formulate the algorithm of multi-memory gradient ascent (MMGA) for asymmetric memory lengths between the players. This discretized algorithm can be implemented even when we do not know the analytical expression of $u^{\eq}(\norm({\bf x}),{\bf y})$.
\begin{algorithm}[H]
    \caption{Discretized MMGA}
    \label{D-MMGA}
    \textbf{Input}: $\eta$, $\gamma$
    \begin{algorithmic}[1]
        \FOR{$t=0,1,2,\cdots$}
        \FOR{$a\in\mc{A}$, $s\in\mc{S}$}
        \STATE ${\bf x}'\gets {\bf x}+\gamma{\bf e}_{a|s}$
        \STATE $\displaystyle \Delta_{a|s}\gets\frac{u^{\eq}(\norm({\bf x}'),{\bf y})-u^{\eq}({\bf x},{\bf y})}{\gamma}$
        \ENDFOR
        \FOR{$a\in\mc{A}$, $s\in\mc{S}$}
        \STATE $x_{a|s}\gets x_{a|s}(1+\eta\Delta_{a|s})$
        \ENDFOR
        \STATE ${\bf x}\gets\norm({\bf x})$
        \ENDFOR
    \end{algorithmic}
\end{algorithm}

In Alg.~\ref{D-MMGA}, the initial states of ${\bf x}$ and ${\bf y}$ are set. In each time step of $t$, X shifts each component of ${\bf x}$ by $\gamma$ from the original strategy ${\bf x}$ and define ${\bf x}'$ (line 3). Here, ${\bf e}_{a|s}$ is defined as the unit vector for the direction of $x_{a|s}$. Furthermore, we normalize ${\bf x}'$ for each state as $\norm({\bf x'}):=\{x'_{a|s}/(\sum_{a}x'_{a|s})\}_{a,s}$. By comparing $\norm({\bf x}')$ with ${\bf x}$, X gets the gradient $\Delta_{a|s}$ (line 4). Using all of $\Delta_{a|s}$, X updates its strategy $x_{a|s}$ with learning rate $\eta$ (line 7) and normalizes it (line 8). Here, note that the update is weighted by $x_{a|s}$ itself. We also implement Alg.~\ref{D-MMGA} for agent Y similarly. However, because Y has a shorter memory than X, its strategy is updated for $s_{n_{\rm Y}}\in\mc{S}_{n_{\rm Y}}$ instead of $s\in\mc{S}$. Y's action $b\in\mc{B}$ and payoff $v^{\eq}$ can be different from $a\in\mc{A}$ and $u^{\eq}$.
% Figure 03
\begin{figure*}[ht]
    \centering
    \includegraphics[width=0.8\hsize]{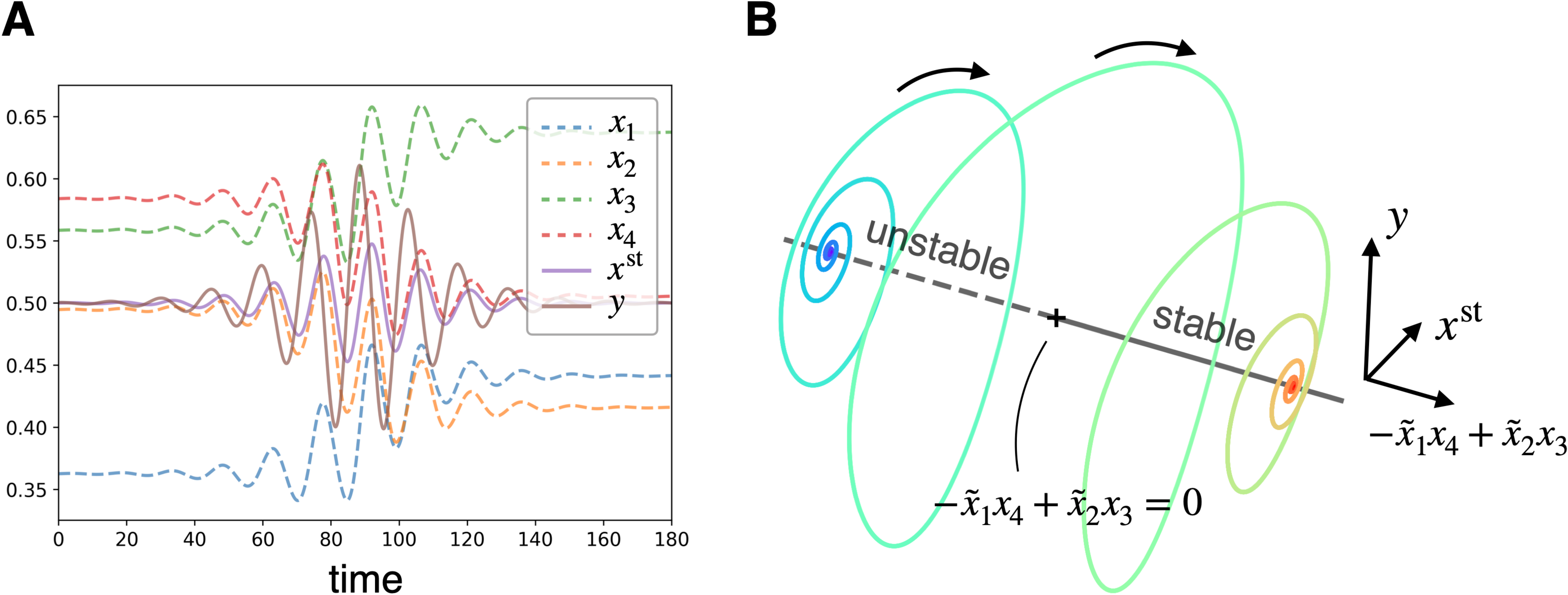}
    \caption{An example of heteroclinic orbit. A: Time series of $\bs{x}=(x_1,x_2,x_3,x_4)$, $x^{\eq}(\bs{x},y)$, and $y$. B: Illustration of trajectories of learning dynamics. The trajectory is plotted in a three-dimensional space consisting of $x^{\eq}(\bs{x},y)$, $y$, and $-\tilde{x}_1x_4+\tilde{x}_2x_3$. Only the solid line and the cross mark (i.e., $-\tilde{x}_1x_4+\tilde{x}_2x_3\ge 0$) is the Nash equilibrium, but the solid and dashed gray lines indicate the states that correspond to the original Nash equilibrium, i.e., $(x^{\eq}(\bs{x},y),y)=(x^{\rm o},y^{\rm o})$. Note that this line should be a three-dimensional manifold in the five-dimensional space of $\bs{x}$ and $y$ in practice. The solid (resp. dashed) line satisfies the stable condition $-\tilde{x}_1x_4+\tilde{x}_2x_3> 0$ (resp. $<0$). The trajectory is plotted from blue (time is $0$) to red (time is $180$) by the time series in panel A.
    }
    \label{F03}
\end{figure*}

If the analytical expression of $u^{\eq}({\bf x}, {\bf y})$ and $v^{\eq}({\bf x}, {\bf y})$ are known, we can use the following dynamics (continualized MMGA~\cite{fujimoto2023learning}) instead of Alg.~\ref{D-MMGA} by taking the limit of $\gamma\to 0$ and $\eta\to 0$;
\begin{align}
    \dot{x}_{a|s}({\bf x}, {\bf y})&=x_{a|s}\frac{\partial}{\partial x_{a|s}}u^{\eq}(\norm({\bf x}),{\bf y}),
    \label{C-MMGA1}\\
    \dot{y}_{b|s_{n_{\rm Y}}}({\bf x}, {\bf y})&=y_{b|s_{n_{\rm Y}}}\frac{\partial}{\partial y_{b|s_{n_{\rm Y}}}}v^{\eq}({\bf x},\norm({\bf y})).
    \label{C-MMGA2}
\end{align}
These continuous-time dynamics of the MMGA algorithm are useful for the theoretical treatment as they allow for classical stability analysis and numerical integration.

\section{Theoretical Results}
In this section, we analyze learning dynamics under asymmetric memory lengths between the players. To simplify theoretical treatments, we first establish several assumptions for the games, such as payoff matrix, action number, and memory length. Next, we derive the Nash equilibrium in these games and compare this equilibrium with the original Nash equilibrium in games without memory. Finally, we analyze the dynamics of learning. We prove that the fixed points of learning dynamics are divided into stable and unstable ones. Thus, heteroclinic dynamics diverging from unstable fixed points and converging to stable ones are drawn. In this section, we present only the proof sketch for each theorem, while detailed proofs can be found in Technical Appendix~A.

\subsection{Assumptions}
First, we consider two-action zero-sum games as follows.

\begin{assumption}[Two-action zero-sum game]
\label{asm_Two-action}
We denote the two actions of two agents as $\mc{A}=\{a_1,a_2\}$ and $\mc{B}=\{b_1,b_2\}$. There are four states $\mc{S}_1=\mc{A}\times\mc{B}=\{a_1b_1,a_1b_2,a_2b_1,a_2b_2\}$. If each state occurs, X receives the payoff corresponding to $\bs{u}=(u_1,\ldots,u_4):=(u_{a_1b_1},u_{a_1b_2},u_{a_2b_1},u_{a_2b_2})$. Zero-sum games assume that Y's payoff is given by $v_{ab}=-u_{ab}$ for all $a\in\mc{A}$ and $b\in\mc{B}$. To ensure a nontrivial game, we assume that $u_1$ and $u_4$ are both larger than $u_2$ and $u_3$.
\end{assumption}

In Asm.~\ref{asm_Two-action}, we consider a specific class of zero-sum games where the Nash equilibrium exists in the interior of the strategy spaces of two players. This is because if the Nash equilibrium is on the boundary of the strategy spaces, a dominant strategy exists and learning dynamics are trivial. An example that satisfies Asm.~\ref{asm_Two-action} is matching-pennies games ($u_1=u_4=+1$ and $u_2=u_3=-1$). Indeed, the Nash equilibrium is defined as follows.

\begin{definition}[Original Nash equilibrium in two-action zero-sum normal-form game]
\label{def_Nash}
Under Asm.~\ref{asm_Two-action} (two-action zero-sum game), we consider $(n_{\rm X},n_{\rm Y})=(0,0)$ (no memory = normal-form game). Then, the mixed strategy of X is given by a single variable $x:=x_{a_1}\in(0,1)$. Similarly, Y's strategy is given by $y:=y_{b_1}\in(0,1)$. Then, the Nash equilibrium $(x^{\rm o},y^{\rm o})$ and the payoffs in the equilibrium $(u^{\rm o},v^{\rm o})$ are given by
\begin{align}
    &x^{\rm o}=\frac{-u_3+u_4}{u_1-u_2-u_3+u_4},\ y^{\rm o}=\frac{-u_2+u_4}{u_1-u_2-u_3+u_4},\\
    &u^{\rm o}=\frac{u_1u_4-u_2u_3}{u_1-u_2-u_3+u_4},\ v^{\rm o}=-u^{\rm o}.
\end{align}
\end{definition}

The Nash equilibrium in no-memory games has been known for a long time~\cite{nash1950equilibrium}. On the other hand, when agents with memories are considered, the region of the Nash equilibrium is generally extended~\cite{fudenberg2009folk}. Therefore, we call such a no-memory Nash equilibrium $(x^{\rm o},y^{\rm o})$ as the ``original'' Nash equilibrium for distinction.

We now introduce a specific class of with-memory games under Asm.~\ref{asm_Two-action} and define notations.

\begin{definition}[One-memory and zero-memory strategies and vector notation]
\label{def_One-memory}
Under Asm.~\ref{asm_Two-action}, consider $(n_{\rm X},n_{\rm Y})=(1,0)$. Because a constraint $x_{a_1|s}+x_{a_2|s}=1$ holds for all $s\in\mc{S}=\{a_1b_1,a_1b_2,a_2b_1,a_2b_2\}$, X's strategy is determined by a four-variable vector $\bs{x}=(x_1,\ldots,x_4):=(x_{a_1|a_1b_1}, x_{a_1|a_1b_2}, x_{a_1|a_2b_1}, x_{a_1|a_2b_2})\in(0,1)^{4}$. On the other hand, Y's strategy is determined by $y:=y_{b_1}\in(0,1)$. We also use the vector notation of the stationary state $\bs{p}^{\eq}=(p_1^{\eq},\ldots,p_4^{\eq}):=(p_{a_1b_1}^{\eq},p_{a_1b_2}^{\eq},p_{a_2b_1}^{\eq},p_{a_2b_2}^{\eq})$. Expected payoffs in the stationary state are also described as $u^{\eq}(\bs{x},y):=\bs{p}^{\eq}(\bs{x},y)\cdot\bs{u}$ and $v^{\eq}(\bs{x},y)=-u^{\eq}(\bs{x},y)$.
\end{definition}

\subsection{Analysis of Nash Equilibrium}
First, we provide an important theorem characterizing games between one-memory and zero-memory agents. In the following, we define $\tilde{\mc{X}}:=1-\mc{X}$ for any function or variable $\mc{X}$.

\begin{theorem}[Stationary state]
\label{thm_Memory-less}
Under Def.~\ref{def_One-memory}, the stationary state can be described as $\bs{p}^{\eq}(\bs{x},y)=(x^{\eq},\tilde{x}^{\eq})\otimes(y,\tilde{y}):=(x^{\eq}y,x^{\eq}\tilde{y},\tilde{x}^{\eq}y,\tilde{x}^{\eq}\tilde{y})$. Here, $x^{\eq}$ is called X's ``marginalized'' strategy, a function of $(\bs{x},y)$;
\begin{align}
    x^{\eq}(\bs{x},y)=\frac{x_3y+x_4\tilde{y}}{\tilde{x}_1y+\tilde{x}_2\tilde{y}+x_3y+x_4\tilde{y}}.
\end{align}
\end{theorem}

\noindent {\it Proof Sketch.} We consider the stationary state condition $p_{s'}^{\eq}=\sum_{s}M_{s's}p_{s}^{\eq}$. Because Y uses zero-memory strategies, we derive $p_1^{\eq}+p_3^{\eq}=y$, $p_2^{\eq}+p_4^{\eq}=\tilde{y}$, and $p_1^{\eq}/p_2^{\eq}=p_3^{\eq}/p_4^{\eq}$. These three equations show that the stationary state is described as $\bs{p}^{\eq}=(x^{\eq},\tilde{x}^{\eq})\otimes(y,\tilde{y})$ with a function $x^{\eq}(\bs{x},y)$. By substituting this equation for the stationary state condition, we obtain the mathematical expression of $x^{\eq}(\bs{x},y)$. \qed

Thm.~\ref{thm_Memory-less} shows that in the stationary state, how each action is chosen by any one-memory strategy $\bs{x}\in(0,1)^4$ can be given by a zero-memory strategy $x=x^{\eq}(\bs{x},y)$. Here, because $x^{\eq}(\bs{x},y)$ is obtained by compressing $\bs{x}$ and $y$, we call it a ``marginalized'' strategy (following the terminology used in~\citet{press2012iterated}). The representation of the marginalized strategy is as if X is using the no-memory strategy in the stationary state, but this is because opponent Y has no memory. However, note that this marginalized strategy $x^{\eq}(\bs{x},y)$ is not a variable but a function that changes depending on the other's strategy $y$. Based on Thm.~\ref{thm_Memory-less}, we obtain the Nash equilibrium as follows.

\begin{theorem}[With-memory Nash equilibrium]
\label{thm_Nash}
In with-memory games under Def.~\ref{def_One-memory}, $(\bs{x},y)=(\bs{x}^*,y^*)$ is the with-memory Nash equilibrium if and only if $x^{\eq}(\bs{x}^*,y^*)=x^{\rm o}$, $y^*=y^{\rm o}$, and $-\tilde{x}_1^*x_4^*+\tilde{x}_2^*x_3^*\ge 0$ are satisfied.
\end{theorem}

\noindent{\it Proof Sketch.} We consider the extreme value conditions of X and Y. Using the notation of $\mc{F}|_{*}:=\mc{F}|_{(\bs{x},y)=(\bs{x}^*,y^*)}$ for any functions $\mc{F}$, we derive $\partial u^{\eq}/\partial x_i|_{*}=0\Leftrightarrow y^*=y^{\rm o}$ for all $i$. Here, when $y=y^{\rm o}$, $u^{\eq}$ is always constant independent of $\bs{x}$. We also derive $\partial v^{\eq}/\partial y|_{*}=0\Leftrightarrow x^{\eq}(\bs{x}^*,y^*)=x^{\rm o}$. Furthermore, the concavity condition of $v^{\eq}$ for $y$ is given by
\begin{align}
    \frac{\partial^2 v^{\eq}(\bs{x}^*,y)}{\partial y^2}\le 0\ \ &\Leftrightarrow\ \ \frac{\partial x^{\eq}(\bs{x}^*,y)}{\partial y}\ge 0
    \nonumber\\
    &\Leftrightarrow\ \ -\tilde{x}_1^*x_4^*+\tilde{x}_2^*x_3^*\ge 0,
    \label{equivalence}
\end{align}
completing the proof. \qed

Let us discuss this with-memory Nash equilibrium, which is simply called Nash equilibrium below. First, $(\bs{x}^*,y^*)$ is a three-dimensional manifold in the space of $(\bs{x},y)$. In the space of $(x^{\eq}(\bs{x},y),y)$, however, $(x^{\eq}(\bs{x}^*,y^*),y^*)$ is unique and corresponds to the original Nash equilibrium $(x^{\rm o},y^{\rm o})$. Here, note that not all of $(\bs{x},y)$ such that $(x^{\eq}(\bs{x},y),y)=(x^{\rm o},y^{\rm o})$ are the Nash equilibrium $(\bs{x}^*,y^*)$. This is because one more condition $-\tilde{x}_1^*x_4^*+\tilde{x}_2^*x_3^*\ge 0$ is necessary for the Nash equilibrium.
% Figure 04
\begin{figure*}[ht]
    \centering
    \includegraphics[width=0.8\hsize]{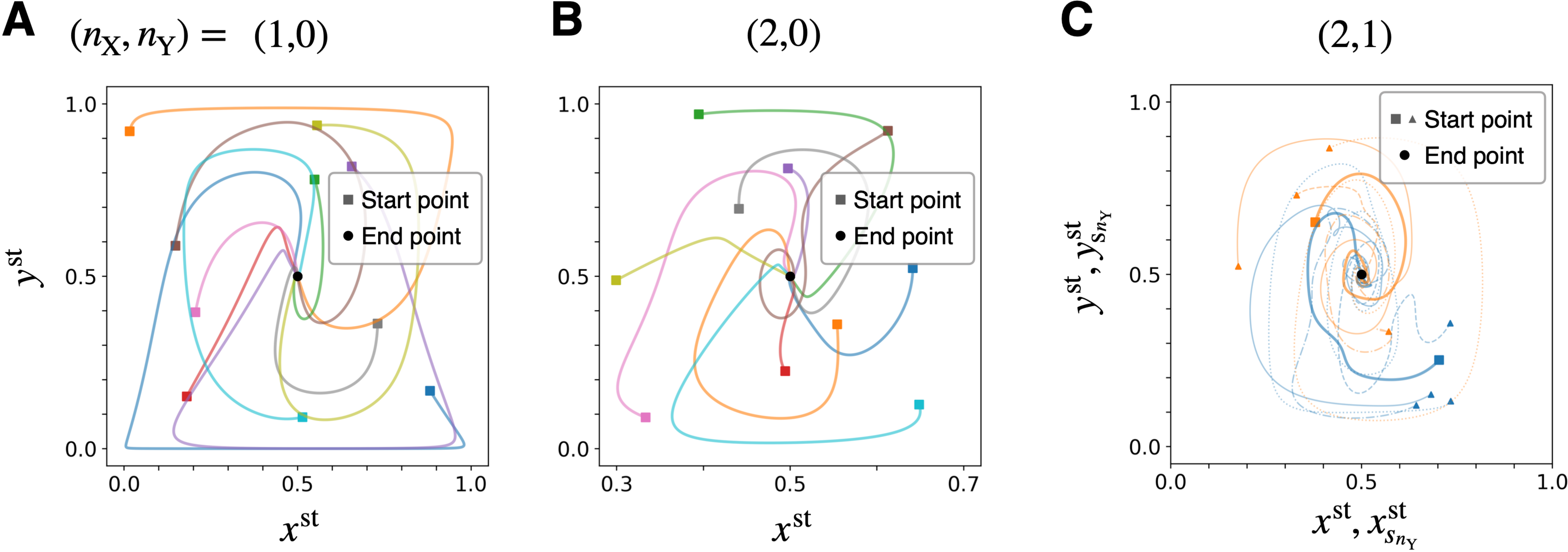}
    \caption{Learning dynamics for various asymmetric numbers of memories by removing Def.~\ref{def_One-memory}. Panels A and B show the cases of $(n_{\rm X},n_{\rm Y})=(1,0)$ and $(2,0)$, respectively. All the trajectories from various $10$ random initial states (represented by different colors) converge to the same Nash equilibrium. Panel C illustrates the case of $(n_{\rm X},n_{\rm Y})=(2,1)$. The horizontal and vertical axes indicate $x^{\eq}$ and $y^{\eq}$ for the wide thick lines, and X's marginalized strategy $x_{s_{n_{\rm Y}}}^{\eq}$ and Y's strategy $y_{s_{n_{\rm Y}}}^{\eq}$ for the other thin lines. The solid, dashed, dot, and dash-dot lines of the thin ones represent the cases of $s_{n_{\rm Y}}(=s_1)=a_1b_1$, $a_1b_2$, $a_2b_1$, and $a_2b_2$, respectively. Both samples (represented by blue and orange) are observed to converge to the Nash equilibrium.
    }
    \label{F04}
\end{figure*}

We also discuss this condition $-\tilde{x}_1^*x_4^*+\tilde{x}_2^*x_3^*\ge 0$. First, recall that X uses four variables $\bs{x}$ to construct a function $x^{\eq}(\bs{x},y)$. As shown in Thm.~\ref{thm_Memory-less}, this function means the probability that X chooses action $a_1$ in the stationary state. Practically, however, X can change this probability depending on the other's strategy $y$ because it has memory. $-\tilde{x}_1^*x_4^*+\tilde{x}_2^*x_3^*\ge 0$ is satisfied when $x_1^*$ and $x_3^*$ are large while $x_2^*$ and $x_4^*$ are small. In other words, X tends to use $a_1$ (resp. $a_2$) in response to $b_1$ (resp. $b_2$), meaning that X, in the next round, tries to use the advantageous action in response to the opponent's action. Briefly said, X exploits Y's payoff if Y biasedly chooses an action. Thus, it is best for Y to use its minimax strategy $y=y^*$. Indeed, $-\tilde{x}_1^*x_4^*+\tilde{x}_2^*x_3^*\ge 0$ is equivalent to $\partial^2 v^{\eq}(\bs{x}^*,y)/\partial y^2\le 0$, meaning that Y's function is concave. Y cannot increase its own payoff even if it uses other strategies $y\neq y^*$. Thus, this condition is necessary for the Nash equilibrium.

\subsection{Analysis of Learning Dynamics}
Having discussed the appearance of a Nash equilibrium in a game with asymmetric memory, our attention will now be directed towards the dynamics of the game. 
Specifically, We now analyze the dynamics of Eqs.~\eqref{C-MMGA1} and \eqref{C-MMGA2} around the equilibrium. 
Our first finding is that the fixed points of the learning dynamics correspond to the original Nash equilibrium.

\begin{theorem}[Fixed points of learning dynamics]
\label{thm_Fixed}
Under Def.~\ref{def_One-memory}, all the fixed points of learning dynamics are given by $(x^{\eq}(\bs{x},y),y)=(x^{\rm o},y^{\rm o})$.
\end{theorem}

\noindent{\it Proof Sketch.} Under Def.~\ref{def_One-memory}, Eqs.~\eqref{C-MMGA1} and \eqref{C-MMGA2} are calculated as
\begin{align}
    \dot{x}_i&=x_i\tilde{x}_i(u_1-u_2-u_3+u_4)(y-y^{\rm o})\frac{\partial x^{\eq}(\bs{x},y)}{\partial x_i},\\
    \dot{y}&=-y\tilde{y}(u_1-u_2-u_3+u_4)
    \nonumber\\
    &\hspace{0.3cm}\times \left\{(y-y^{\rm o})\frac{\partial x^{\eq}(\bs{x},y)}{\partial y}+(x^{\eq}(\bs{x},y)-x^{\rm o})\right\}.
\end{align}
We can calculate $(\dot{\bs{x}},\dot{y})=(\bs{0},0)\Leftrightarrow (x^{\eq}(\bs{x},y),y)=(x^{\rm o},y^{\rm o})$. \qed

Next, the following theorem and corollary specify whether dynamics converge to or diverge from these fixed points.

\begin{theorem}[Local convergence to Nash equilibrium]
\label{thm_Local}
Under Def.~\ref{def_One-memory}, the learning dynamics are locally asymptotically stable for each Nash equilibrium $(\bs{x}^*,y^*)$ such that $-\tilde{x}_1^*x_4^*+\tilde{x}_2^*x_3^* > 0$.
\end{theorem}

\noindent{\it Proof Sketch.} We consider a linear stability analysis in the neighborhoods of the obtained fixed points. Let $\bs{J}$ denote the Jacobian matrix for learning dynamics with $\bs{z}=(\bs{x},y)$ and $J_{ij}:=\partial \dot{z}_i/ \partial z_j|_{*}$. Each fixed point is locally stable when the maximum eigenvalue of the Jacobian except for $0$ (denoted as $\lambda_1$) is negative. We derive $\lambda_1< 0 \Leftrightarrow -\tilde{x}_1^*x_4^*+\tilde{x}_2^*x_3^*> 0$, completing the proof. \qed

\begin{corollary}[Divergence from fixed points]
\label{cor_Divergence}
Under Def.~\ref{def_One-memory}, if strategies $\bs{x}$ and $y$ satisfy $(x^{\eq}(\bs{x},y),y)=(x^{\rm o},y^{\rm o})$, the strategies are fixed points of learning dynamics even in the region of $-\tilde{x}_1x_4+\tilde{x}_2x_3< 0$. However, the strategies are unstable, and learning dynamics diverge from there.
\end{corollary}
% Figure 05
\begin{figure*}[ht]
    \centering
    \includegraphics[width=0.55\hsize]{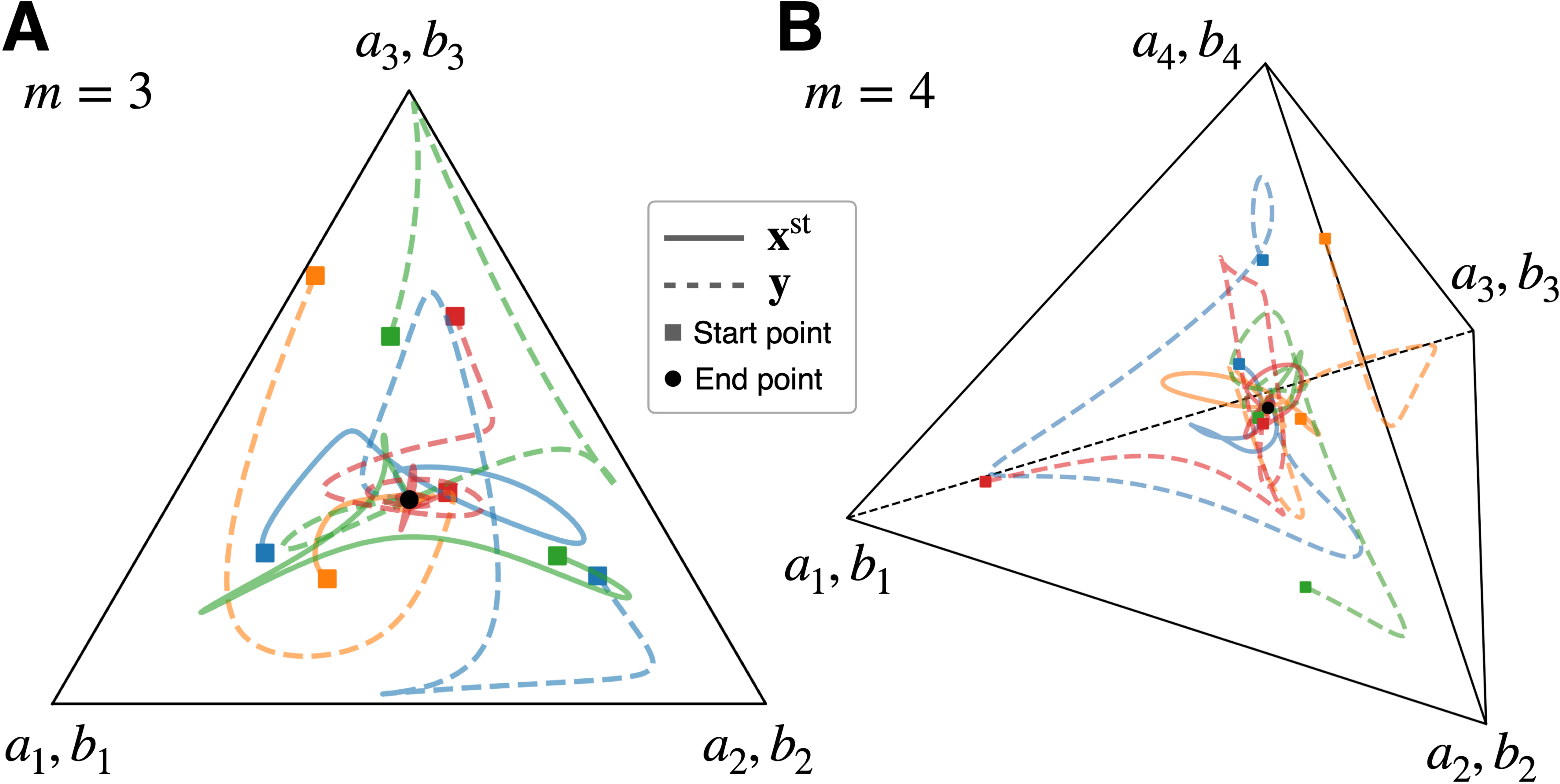}
    \caption{Learning dynamics for various numbers of actions are illustrated in two cases: (A) a rock-paper-scissors game ($m=3$), and (B) an extended rock-paper-scissors game with four actions ($m=4$). The dynamics of ${\bf x}^{\eq}\in\Delta^{m-1}$ and ${\bf y}\in\Delta^{m-1}$ are plotted in each panel. The vertices of simplexes correspond to pure strategies for choosing actions $a_1$, $\cdots$, or $a_m$ in ${\bf x}^{\eq}$ or $b_1$, $\cdots$, or $b_m$ in ${\bf y}$. In both cases, the samples of $4$ random initial conditions (indicated by different colors) converge to the Nash equilibrium.
    }
    \label{F05}
\end{figure*}

Let us explain the property of learning dynamics from Thms.~\ref{thm_Fixed} and \ref{thm_Local}. First, Thm.~\ref{thm_Fixed} means that the fixed points of learning dynamics correspond to the original Nash equilibrium. However, learning dynamics do not converge to all of these fixed points. Thm.~\ref{thm_Local} shows that learning can converge to only roughly half of the fixed points where the condition $-\tilde{x}_1x_4+\tilde{x}_2x_3>0$ should be satisfied. Because the equivalence holds when all the inequalities become equalities in Eq.~\eqref{equivalence}, this condition is equivalent to the strict concavity of Y's utility function $v^{\eq}(\bs{x},y)$.

\subsection{Visualization of Heteroclinic Dynamics}
Fig.~\ref{F03} visualizes a heteroclinic orbit of learning dynamics. Panel A shows the time series of learning dynamics which is based on the continuous-time algorithm, i.e., the Runge-Kutta fourth-order method of Eqs.~\eqref{C-MMGA1} and \eqref{C-MMGA2} with the step-size of $2\times 10^{-2}$. To facilitate interpretation, Panel B shows the trajectory of the time series of five-dimensional learning dynamics of $(\bs{x},y)$ in an appropriate three-dimensional space $(x^{\eq},y,-\tilde{x}_1x_4+\tilde{x}_2x_3)$. In panel B, all the gray solid and dashed lines correspond to the original Nash equilibrium, i.e., $(x^{\eq},y)=(x^{\rm o},y^{\rm o})$ and to the fixed points of learning dynamics. In particular, the solid (resp. dashed) line shows stable (resp. unstable) fixed points of learning dynamics. Note that these one-dimensional lines represent a three-dimensional manifold in the five-dimensional space consisting of $(\bs{x},y)$. Now, we can see a heteroclinic orbit diverging from an unstable fixed point and converging to a stable one. Panel A illustrates an increase in $x_1$ and $x_3$, while $x_2$ and $x_4$ decrease through learning, indicating that X is learning to exploit Y. Consequently, $-\tilde{x}_1x_4+\tilde{x}_2x_3$ increases, leading to a concave utility function for Y with respect to $y$.
%%%%%%% Figure 03 %%%%%%%

\section{Experimental Results}
So far, our theoretical analyses of learning dynamics have revealed the existence of heteroclinic orbits that converge to stable fixed points. In addition, these fixed points are included in the original Nash equilibrium of games without memory. In the following, we numerically confirm that learning dynamics converge to the original Nash equilibrium in various initial states, action numbers, and memory lengths. All the following experimental results are based on the discrete-time algorithm, i.e., Alg.~\ref{D-MMGA}. In the following, we set the inputs $\eta=10^{-3}$ and $\gamma=10^{-6}$ in Alg.~\ref{D-MMGA}. In computing gradients of payoff, i.e., $\Delta_{a|s}$, we calculate the equilibrium payoff accurately enough; The analytical solution $u^{\eq}$ and computational solution $\hat{u}^{\eq}$ satisfy $|u^{\eq}-\hat{u}^{\eq}|\le 10^{-9}$.

\paragraph{Dynamics for Various Memory Lengths.}
Fig.~\ref{F04} shows that the learning dynamics converge to the Nash equilibrium in various numbers of memories. Panel A is for $(n_{\rm X},n_{\rm Y})=(1,0)$, B is for $(2,0)$, and C is for $(2,1)$. In Panel B, because $n_{\rm X}=2$, the number of variables that are necessary to describe X's strategy $\bs{x}$ is larger than those in Def.~\ref{def_One-memory}. However, just like Thm.~\ref{thm_Memory-less} shows, $\bs{x}$ construct the marginalized strategy of one function $x^{\eq}(\bs{x},y)$ in the stationary state. Furthermore, the equilibrium corresponds to the original Nash equilibrium, i.e., $(x^{\eq}(\bs{x},y),y)=(x^{\rm o},y^{\rm o})$. Notably, Panel C of $(n_{\rm X},n_{\rm Y})=(2,1)$ is interesting, where both the players memorize previous states as different from both A and B. X uses two-memory strategies given by sixteen variables $\bs{x}\in(0,1)^{16}$, while Y uses one-memory strategies given by four variables $\bs{y}\in(0,1)^{4}$. Then, $\bs{x}$ construct the marginalized strategy of four functions and X behaves as if X employs one-memory strategies. In conclusion, learning dynamics converge to the Nash equilibrium in games between one-memory strategies~\cite{fujimoto2023learning}.
%%%%%%% Figure 04 %%%%%%%

\paragraph{Dynamics for Various Action Numbers.}
Next, we remove the assumption of two-action games (Asm.~\ref{asm_Two-action}). Fig.~\ref{F05} demonstrates that learning dynamics converge in various numbers of actions, where Panel A considers a Rock-Paper-Scissors game ($m=3$), while B considers an Extended Rock-Paper-Scissors game ($m=4$) with $(n_{\rm X}, n_{\rm Y})=(1,0)$. From these panels, we can observe that learning dynamics converge in these games as in the two-action games. We now discuss where learning dynamics converge. Just like the proof in Thm.~\ref{thm_Memory-less}, we define the probability that X chooses action $a$ in the stationary state under $m$-action games as $x_a^{\eq}({\bf x},{\bf y}):=\sum_{s\in\mc{S}}x_{a|s}p_{s}^{\eq}({\bf x},{\bf y})$. We also define the set of $x_a^{\eq}({\bf x},{\bf y})$ as ${\bf x}^{\eq}({\bf x},{\bf y}):=\{x_a^{\eq}({\bf x},{\bf y})\;|\;a\in\mc{A}\}$. Fig.~\ref{F05} shows the last-iterate convergence to the Nash equilibrium, i.e., $({\bf x}^{\eq}({\bf x},{\bf y}),{\bf y})=({\bf x}^{\rm o},{\bf y}^{\rm o})$. The mechanism for this convergence is similar to that in two-action games. Indeed, even in $m$-action zero-sum games, the player with $1$-memory endows a strict concavity to the utility function of the $0$-memory player (see the Technical Appendix~C for a detailed explanation).
%%%%%%% Figure 05 %%%%%%%

\paragraph{Experimental Results with Many Samples.} In order to ensure the reliability of our results, we have simulated the learning dynamics with a large number of samples with random initial strategies. In all the samples, we observed convergence to the Nash equilibrium. Refer to the Technical Appendix~B for detailed data.

\section{Discussion}
This study explores the dynamics of learning in games where agents possess different memory capacities, both theoretically and experimentally. 
In our theoretical contributions, we assumed that two agents with memory lengths of one and zero play two-action zero-sum games. Under this assumption, we proved that the original Nash equilibrium in games without memory is divided into stable (Thm.~\ref{thm_Local}) and unstable fixed points (Cor.~\ref{cor_Divergence}) for learning. Here, heteroclinic orbits are discovered in learning in games with memory asymmetry; their strategies diverge from these unstable fixed points and converge to stable ones. In our experimental contributions, we relaxed the assumption on memory lengths and the number of actions, and then we demonstrated that memory asymmetry triggers this convergence to the original Nash equilibrium across a wider range of action numbers and memory lengths. Considering that dynamics show cycling behavior in games without memory and divergence in games with memory symmetry, this convergence is a nontrivial phenomenon.

Note that we have discovered new insights into computing equilibria, with memory asymmetry serving as one example. Convergent algorithms and dynamics, often referred to as last-iterate convergence, are extensively studied within the community of learning in games. The focus is mainly on computationally efficient algorithms with faster rates of convergence or global convergence guarantees~\cite{daskalakis2019last, golowich2020tight, mertikopoulos2019optimistic, wei2021linear, lei2021last, nguyen2021last, anagnostides2022last, abe2022mutation}. In the majority of these studies, convergence is ensured through modifications to the learning algorithm. A typical example is to incorporate ``optimism'' symmetrically into discretized gradient ascent. Each agent then refines its strategy by naively predicting the opponent's strategy update from past data~\cite{daskalakis2019last}. Our method to achieve convergence differs significantly. We introduced memory asymmetry to alter the strategy spaces of one or both agents. By leveraging memory asymmetry, an agent with a longer memory learns to exploit the opponent's payoff. Consequently, the opponent agent with a shorter memory achieves a strictly concave utility function, leading to convergence. To the best of our knowledge, this is the first study to induce convergence through such asymmetry of the strategy spaces. Our findings may open up new possibilities for achieving convergence in learning games and inspire further research in this field.

\section{Acknowledgements}
Y.F. is supported by JSPS KAKENHI Grant No. 21J01393. K.Ariu is supported by JSPS KAKENHI Grant No.~23K19986.

%\bibliography{aaai24_Fujimoto.bib}

\begin{thebibliography}{36}
\providecommand{\natexlab}[1]{#1}

\bibitem[{Abe, Sakamoto, and Iwasaki(2022)}]{abe2022mutation}
Abe, K.; Sakamoto, M.; and Iwasaki, A. 2022.
\newblock Mutation-Driven Follow the Regularized Leader for Last-Iterate
  Convergence in Zero-Sum Games.
\newblock In \emph{UAI}, 1--10.

\bibitem[{Anagnostides et~al.(2022)Anagnostides, Panageas, Farina, and
  Sandholm}]{anagnostides2022last}
Anagnostides, I.; Panageas, I.; Farina, G.; and Sandholm, T. 2022.
\newblock On last-iterate convergence beyond zero-sum games.
\newblock In \emph{ICML}, 536--581.

\bibitem[{Axelrod and Hamilton(1981)}]{axelrod1981evolution}
Axelrod, R.; and Hamilton, W.~D. 1981.
\newblock The evolution of cooperation.
\newblock \emph{Science}, 211(4489): 1390--1396.

\bibitem[{Baek et~al.(2016)Baek, Jeong, Hilbe, and Nowak}]{baek2016comparing}
Baek, S.~K.; Jeong, H.-C.; Hilbe, C.; and Nowak, M.~A. 2016.
\newblock Comparing reactive and memory-one strategies of direct reciprocity.
\newblock \emph{Scientific reports}, 6(1): 25676.

\bibitem[{Barfuss(2020)}]{barfuss2020reinforcement}
Barfuss, W. 2020.
\newblock Reinforcement learning dynamics in the infinite memory limit.
\newblock In \emph{AAMAS}, 1768--1770.

\bibitem[{Barfuss, Donges, and Kurths(2019)}]{barfuss2019deterministic}
Barfuss, W.; Donges, J.~F.; and Kurths, J. 2019.
\newblock Deterministic limit of temporal difference reinforcement learning for
  stochastic games.
\newblock \emph{Physical Review E}, 99(4): 043305.

\bibitem[{B{\"o}rgers and Sarin(1997)}]{borgers1997learning}
B{\"o}rgers, T.; and Sarin, R. 1997.
\newblock Learning through reinforcement and replicator dynamics.
\newblock \emph{Journal of Economic Theory}, 77(1): 1--14.

\bibitem[{Bowling(2004)}]{bowling2004convergence}
Bowling, M. 2004.
\newblock Convergence and no-regret in multiagent learning.
\newblock In \emph{NeurIPS}, 209--216.

\bibitem[{Bowling and Veloso(2002)}]{bowling2002multiagent}
Bowling, M.; and Veloso, M. 2002.
\newblock Multiagent learning using a variable learning rate.
\newblock \emph{Artificial Intelligence}, 136(2): 215--250.

\bibitem[{Daskalakis and Panageas(2019)}]{daskalakis2019last}
Daskalakis, C.; and Panageas, I. 2019.
\newblock Last-iterate convergence: Zero-sum games and constrained min-max
  optimization.
\newblock In \emph{ITCS}, 27:1--27:18.

\bibitem[{Fudenberg and Levine(1998)}]{fudenberg1998theory}
Fudenberg, D.; and Levine, D.~K. 1998.
\newblock \emph{The theory of learning in games}, volume~2.
\newblock MIT press.

\bibitem[{Fudenberg and Maskin(2009)}]{fudenberg2009folk}
Fudenberg, D.; and Maskin, E. 2009.
\newblock The folk theorem in repeated games with discounting or with
  incomplete information.
\newblock In \emph{A long-run collaboration on long-run games}, 209--230. World
  Scientific.

\bibitem[{Fujimoto, Ariu, and Abe(2023)}]{fujimoto2023learning}
Fujimoto, Y.; Ariu, K.; and Abe, K. 2023.
\newblock Learning in Multi-Memory Games Triggers Complex Dynamics Diverging
  from Nash Equilibrium.
\newblock In \emph{IJCAI}.

\bibitem[{Fujimoto and Kaneko(2019)}]{fujimoto2019emergence}
Fujimoto, Y.; and Kaneko, K. 2019.
\newblock Emergence of exploitation as symmetry breaking in iterated prisoner's
  dilemma.
\newblock \emph{Physical Review Research}, 1(3): 033077.

\bibitem[{Fujimoto and Kaneko(2021)}]{fujimoto2021exploitation}
Fujimoto, Y.; and Kaneko, K. 2021.
\newblock Exploitation by asymmetry of information reference in coevolutionary
  learning in prisoner's dilemma game.
\newblock \emph{Journal of Physics: Complexity}, 2(4): 045007.

\bibitem[{Golowich, Pattathil, and Daskalakis(2020)}]{golowich2020tight}
Golowich, N.; Pattathil, S.; and Daskalakis, C. 2020.
\newblock Tight last-iterate convergence rates for no-regret learning in
  multi-player games.
\newblock In \emph{NeurIPS}, 20766--20778.

\bibitem[{Hilbe et~al.(2017)Hilbe, Martinez-Vaquero, Chatterjee, and
  Nowak}]{hilbe2017memory}
Hilbe, C.; Martinez-Vaquero, L.~A.; Chatterjee, K.; and Nowak, M.~A. 2017.
\newblock Memory-n strategies of direct reciprocity.
\newblock \emph{Proceedings of the National Academy of Sciences}, 114(18):
  4715--4720.

\bibitem[{Hofbauer, Sigmund et~al.(1998)}]{hofbauer1998evolutionary}
Hofbauer, J.; Sigmund, K.; et~al. 1998.
\newblock \emph{Evolutionary games and population dynamics}.
\newblock Cambridge university press.

\bibitem[{Lei et~al.(2021)Lei, Nagarajan, Panageas et~al.}]{lei2021last}
Lei, Q.; Nagarajan, S.~G.; Panageas, I.; et~al. 2021.
\newblock Last iterate convergence in no-regret learning: constrained min-max
  optimization for convex-concave landscapes.
\newblock In \emph{AISTATS}, 1441--1449.

\bibitem[{Mertikopoulos et~al.(2019)Mertikopoulos, Lecouat, Zenati, Foo,
  Chandrasekhar, and Piliouras}]{mertikopoulos2019optimistic}
Mertikopoulos, P.; Lecouat, B.; Zenati, H.; Foo, C.-S.; Chandrasekhar, V.; and
  Piliouras, G. 2019.
\newblock Optimistic mirror descent in saddle-point problems: Going the
  extra(-gradient) mile.
\newblock In \emph{ICLR}.

\bibitem[{Mertikopoulos, Papadimitriou, and
  Piliouras(2018)}]{mertikopoulos2018cycles}
Mertikopoulos, P.; Papadimitriou, C.; and Piliouras, G. 2018.
\newblock Cycles in adversarial regularized learning.
\newblock In \emph{SODA}, 2703--2717.

\bibitem[{Mertikopoulos and Sandholm(2016)}]{mertikopoulos2016learning}
Mertikopoulos, P.; and Sandholm, W.~H. 2016.
\newblock Learning in games via reinforcement and regularization.
\newblock \emph{Mathematics of Operations Research}, 41(4): 1297--1324.

\bibitem[{Meylahn, Janssen et~al.(2022)}]{meylahn2022limiting}
Meylahn, J.~M.; Janssen, L.; et~al. 2022.
\newblock Limiting dynamics for Q-learning with memory one in symmetric
  two-player, two-action games.
\newblock \emph{Complexity}, 2022.

\bibitem[{Murase and Baek(2020)}]{murase2020five}
Murase, Y.; and Baek, S.~K. 2020.
\newblock Five rules for friendly rivalry in direct reciprocity.
\newblock \emph{Scientific reports}, 10(1): 16904.

\bibitem[{Nash~Jr(1950)}]{nash1950equilibrium}
Nash~Jr, J.~F. 1950.
\newblock Equilibrium points in n-person games.
\newblock \emph{Proceedings of the National Academy of Sciences}, 36(1):
  48--49.

\bibitem[{Nguyen et~al.(2021)Nguyen, Zemhoho, Tran-Thanh
  et~al.}]{nguyen2021last}
Nguyen, T.-D.; Zemhoho, A.~B.; Tran-Thanh, L.; et~al. 2021.
\newblock Last round convergence and no-dynamic regret in asymmetric repeated
  games.
\newblock In \emph{ALT}, 553--577.

\bibitem[{Press and Dyson(2012)}]{press2012iterated}
Press, W.~H.; and Dyson, F.~J. 2012.
\newblock Iterated Prisoner’s Dilemma contains strategies that dominate any
  evolutionary opponent.
\newblock \emph{Proceedings of the National Academy of Sciences}, 109(26):
  10409--10413.

\bibitem[{Sandholm and Crites(1996)}]{sandholm1996multiagent}
Sandholm, T.~W.; and Crites, R.~H. 1996.
\newblock Multiagent reinforcement learning in the iterated prisoner's dilemma.
\newblock \emph{Biosystems}, 37(1-2): 147--166.

\bibitem[{Sato, Akiyama, and Farmer(2002)}]{sato2002chaos}
Sato, Y.; Akiyama, E.; and Farmer, J.~D. 2002.
\newblock Chaos in learning a simple two-person game.
\newblock \emph{Proceedings of the National Academy of Sciences}, 99(7):
  4748--4751.

\bibitem[{Schmid et~al.(2022)Schmid, Hilbe, Chatterjee, and
  Nowak}]{schmid2022direct}
Schmid, L.; Hilbe, C.; Chatterjee, K.; and Nowak, M.~A. 2022.
\newblock Direct reciprocity between individuals that use different strategy
  spaces.
\newblock \emph{PLoS Computational Biology}, 18(6): e1010149.

\bibitem[{Singh, Kearns, and Mansour(2000)}]{singh2000nash}
Singh, S.; Kearns, M.~J.; and Mansour, Y. 2000.
\newblock Nash Convergence of Gradient Dynamics in General-Sum Games.
\newblock In \emph{UAI}, 541--548.

\bibitem[{Strogatz(2018)}]{strogatz2018nonlinear}
Strogatz, S.~H. 2018.
\newblock \emph{Nonlinear dynamics and chaos with student solutions manual:
  With applications to physics, biology, chemistry, and engineering}.
\newblock CRC press.

\bibitem[{Ueda(2023)}]{ueda2023memory}
Ueda, M. 2023.
\newblock Memory-two strategies forming symmetric mutual reinforcement learning
  equilibrium in repeated prisoners’ dilemma game.
\newblock \emph{Applied Mathematics and Computation}, 444: 127819.

\bibitem[{Usui and Ueda(2021)}]{usui2021symmetric}
Usui, Y.; and Ueda, M. 2021.
\newblock Symmetric equilibrium of multi-agent reinforcement learning in
  repeated prisoner’s dilemma.
\newblock \emph{Applied Mathematics and Computation}, 409: 126370.

\bibitem[{Wei et~al.(2021)Wei, Lee, Zhang, and Luo}]{wei2021linear}
Wei, C.-Y.; Lee, C.-W.; Zhang, M.; and Luo, H. 2021.
\newblock Linear Last-iterate Convergence in Constrained Saddle-point
  Optimization.
\newblock In \emph{ICLR}.

\bibitem[{Zinkevich(2003)}]{zinkevich2003online}
Zinkevich, M. 2003.
\newblock Online convex programming and generalized infinitesimal gradient
  ascent.
\newblock In \emph{ICML}, 928--936.

\end{thebibliography}

\newpage
\appendix
\onecolumn

\setcounter{secnumdepth}{1}

\renewcommand{\theequation}{A\arabic{equation}}
\setcounter{equation}{0}
\renewcommand{\figurename}{Figure A}
\setcounter{figure}{0}
\renewcommand{\thealgorithm}{A\arabic{algorithm}}
\setcounter{algorithm}{0}

\begin{center}
{\Large {\bf Technical Appendix}}
\end{center}

\section{Proofs} \label{AS01}
\subsection*{Proof of Theorem 1}
Using the stationary state condition $p_{s'}^{\eq}=\sum_{s}M_{s's}p_{s}^{\eq}$ is described for all $s'$ as
\begin{align}
    p_1^{\eq}=x_1yp_1^{\eq}+x_2yp_2^{\eq}+x_3yp_3^{\eq}+x_4yp_4^{\eq},
    \label{st.con.1}\\
    p_2^{\eq}=x_1\tilde{y}p_1^{\eq}+x_2\tilde{y}p_2^{\eq}+x_3\tilde{y}p_3^{\eq}+x_4\tilde{y}p_4^{\eq},
    \label{st.con.2}\\
    p_3^{\eq}=\tilde{x}_1yp_1^{\eq}+\tilde{x}_2yp_2^{\eq}+\tilde{x}_3yp_3^{\eq}+\tilde{x}_4yp_4^{\eq},
    \label{st.con.3}\\
    p_4^{\eq}=\tilde{x}_1\tilde{y}p_1^{\eq}+\tilde{x}_2\tilde{y}p_2^{\eq}+\tilde{x}_3\tilde{y}p_3^{\eq}+\tilde{x}_4\tilde{y}p_4^{\eq},
    \label{st.con.4}
\end{align}
By taking the summation of Eqs.~\eqref{st.con.1} and \eqref{st.con.3} (resp. Eqs.~\eqref{st.con.2} and \eqref{st.con.4}), we trivially obtain $p_1^{\eq}+p_3^{\eq}=y$ (resp. $p_2^{\eq}+p_4^{\eq}=\tilde{y}$). By using Eqs.~\eqref{st.con.1} and \eqref{st.con.2} (resp. Eqs.~\eqref{st.con.3} and \eqref{st.con.4}), we also obtain $p_1^{\eq}/p_2^{\eq}=y/\tilde{y}$ (resp. $p_3^{\eq}/p_4^{\eq}=y/\tilde{y}$). Thus, $p_1^{\eq}/(p_1^{\eq}+p_3^{\eq})=p_2^{\eq}/(p_2^{\eq}+p_4^{\eq})=:x^{\eq}$ holds. From the definition of $x^{\eq}$, the stationary state is described as $\bs{p}^{\eq}=(x^{\eq}y,x^{\eq}\tilde{y},\tilde{x}^{\eq}y,\tilde{x}^{\eq}\tilde{y})=(x^{\eq},\tilde{x}^{\eq})\otimes(y,\tilde{y})$. Moreover, $x^{\eq}$ is obtained as a function of $(\bs{x},y)$ by substituting $\bs{p}^{\eq}=(x^{\eq},\tilde{x}^{\eq})\otimes(y,\tilde{y})$ into Eq.~\eqref{st.con.1};
\begin{align}
    &x^{\eq}y=x_1x^{\eq}y+x_2x^{\eq}\tilde{y}+x_3\tilde{x}^{\eq}y+x_4\tilde{x}^{\eq}\tilde{y}\\
    &\Leftrightarrow x^{\eq}(\bs{x},y)=\frac{x_3y+x_4\tilde{y}}{\tilde{x}_1y+\tilde{x}_2\tilde{y}+x_3y+x_4\tilde{y}}.
\end{align}
\qed

\subsection*{Proof of Theorem 2}
First, we consider the extreme value condition for $u^{\eq}$;
\begin{align}
    \left.\frac{\partial u^{\eq}(\bs{x},y)}{\partial x_i}\right|_{\eqcon}=0&\Leftrightarrow \underbrace{\left.\frac{\partial x^{\eq}(\bs{x},y)}{\partial x_i}\right|_{\eqcon}}_{\neq 0}\left\{(u_1-u_2-u_3+u_4)y^*+(u_2-u_4)\right\}=0,\\
    &\Leftrightarrow y^*=\frac{-u_2+u_4}{u_1-u_2-u_3+u_4}=y^{\rm o}.
    \label{NEcon_1}\\
\end{align}
When Eq.~\eqref{NEcon_1} is satisfied, the payoff of X is always constant independent of $\bs{x}$ as
\begin{align}
    u^{\eq}(\bs{x},y^*)&=\bs{p}^{\eq}(\bs{x},y^*)\cdot\bs{u}\\
    &=x^{\eq}(\bs{x},y^{*})y^{*}u_1+x^{\eq}(\bs{x},y^{*})\tilde{y}^{*}u_2+\tilde{x}^{\eq}(\bs{x},y^{*})y^{*}u_3+\tilde{x}^{\eq}(\bs{x},y^{*})\tilde{y}^{*}u_4\\
    &=x^{\eq}(\bs{x},y^{*})\underbrace{\left\{(u_1-u_2-u_3+u_4)y^{*}+u_2-u_4\right\}}_{=0}+y^{*}(u_3-u_4)+u_4\\
    &=\frac{u_1u_4-u_2u_3}{u_1-u_2-u_3+u_4}\\
    &=u^{\rm o}\ .
\end{align}
Next, we consider the extreme value condition for $v^{\eq}$;
\begin{align}
    \left.\frac{\partial v^{\eq}(\bs{x},y)}{\partial y}\right|_{\eqcon}=0&\Leftrightarrow -\left.\frac{\partial x^{\eq}(\bs{x},y)}{\partial y}\right|_{\eqcon}\underbrace{\left\{(u_1-u_2-u_3+u_4)y^*+u_2-u_4\right\}}_{=0}\nonumber\\
    &\hspace{0.45cm}-\left\{(u_1-u_2-u_3+u_4)x^{\eq}(\bs{x}^*,y^*)+u_3-u_4\right\}=0\\
    &\Leftrightarrow x^{\eq}(\bs{x}^*,y^*)=\frac{-u_3+u_4}{u_1-u_2-u_3+u_4}=x^{\rm o},
    \label{NEcon_3}
\end{align}
In addition, using
\begin{align}
    \frac{\partial x^{\eq}(\bs{x}^*,y)}{\partial y}&=\frac{-\tilde{x}_1^*x_4^*+\tilde{x}_2^*x_3^*}{(\tilde{x}_1^*y+\tilde{x}_2^*\tilde{y}+x_3^*y+x_4^*\tilde{y})^2},\\
    \frac{\partial^2 x^{\eq}(\bs{x}^*,y)}{\partial y^2}&=-2\frac{(-\tilde{x}_1^*x_4^*+\tilde{x}_2^*x_3^*)(\tilde{x}_1^*-\tilde{x}_2^*+x_3^*-x_4^*)}{(\tilde{x}_1^*y+\tilde{x}_2^*\tilde{y}+x_3^*y+x_4^*\tilde{y})^3}=-2\frac{\tilde{x}_1^*-\tilde{x}_2^*+x_3^*-x_4^*}{\tilde{x}_1^*y+\tilde{x}_2^*\tilde{y}+x_3^*y+x_4^*\tilde{y}}\frac{\partial x^{\eq}(\bs{x}^*,y)}{\partial y},
\end{align}
we obtain
\begin{align}
    \frac{\partial^2 v^{\eq}(\bs{x}^*,y)}{\partial y^2}&=-\frac{\partial^2}{\partial y^2}(x^{\eq}(\bs{x}^*,y)yu_1+x^{\eq}(\bs{x}^*,y)\tilde{y}u_2+\tilde{x}^{\eq}(\bs{x}^*,y)yu_3+\tilde{x}^{\eq}(\bs{x}^*,y)\tilde{y}u_4)\\
    &=-(u_1-u_2-u_3+u_4)2\frac{\partial x^{\eq}(\bs{x}^*,y)}{\partial y}-\left\{(u_1-u_2-u_3+u_4)y+u_2-u_4\right\}\frac{\partial^2 x^{\eq}(\bs{x}^*,y)}{\partial y^2}\\
    &=-(u_1-u_2-u_3+u_4)\left\{2\frac{\partial x^{\eq}(\bs{x}^*,y)}{\partial y}+(y-y^*)\frac{\partial^2 x^{\eq}(\bs{x}^*,y)}{\partial y^2}\right\}\\
    &=-(u_1-u_2-u_3+u_4)2\frac{\partial x^{\eq}(\bs{x}^*,y)}{\partial y}\left\{1-(y-y^*)\frac{\tilde{x}_1^*-\tilde{x}_2^*+x_3^*-x_4^*}{\tilde{x}_1^*y+\tilde{x}_2^*\tilde{y}+x_3^*y+x_4^*\tilde{y}}\right\}\\
    &=-(u_1-u_2-u_3+u_4)2\frac{\partial x^{\eq}(\bs{x}^*,y)}{\partial y}\frac{\tilde{x}_1^*y^*+\tilde{x}_2^*\tilde{y}^*+x_3^*y^*+x_4^*\tilde{y}^*}{\tilde{x}_1^*y+\tilde{x}_2^*\tilde{y}+x_3^*y+x_4^*\tilde{y}}.
\end{align}
Since $v^{\eq}(\bs{x}^*,y)$ should be concave for $y$, the condition
\begin{align}
    \frac{\partial^2 v^{\eq}(\bs{x}^*,y)}{\partial y^2}\le 0\ \ \Leftrightarrow\ \ \frac{\partial x^{\eq}(\bs{x}^*,y)}{\partial y}\ge 0\ \ \Leftrightarrow\ \ -\tilde{x}_1^*x_4^*+\tilde{x}_2^*x_3^*\ge 0,
    \label{NEcon_4}
\end{align}
should be satisfied. \qed

\subsection*{Proof of Theorem 3}
We can formulate continualized MMGA under Def.~\ref{def_One-memory};
\begin{align}
    \dot{x}_i&=x_i\tilde{x}_i\frac{\partial u^{\eq}(\bs{x},y)}{\partial x_i}\\
    &=x_i\tilde{x}_i\left\{(u_1-u_2-u_3+u_4)y+(u_2-u_4)\right\}\frac{\partial x^{\eq}(\bs{x},y)}{\partial x_i}\\
    &=x_i\tilde{x}_i(u_1-u_2-u_3+u_4)(y-y^{\rm o})\frac{\partial x^{\eq}(\bs{x},y)}{\partial x_i},\\
    \dot{y}&=-y\tilde{y}\frac{\partial u^{\eq}(\bs{x},y)}{\partial y}\\
    &=-y\tilde{y}\left\{\frac{\partial x^{\eq}(\bs{x},y)}{\partial y}\left\{(u_1-u_2-u_3+u_4)y+(u_2-u_4)\right\}+\left\{(u_1-u_2-u_3+u_4)x^{\eq}(\bs{x},y)+(u_3-u_4)\right\}\right\}\\
    &=-y\tilde{y}(u_1-u_2-u_3+u_4)\left\{(y-y^{\rm o})\frac{\partial x^{\eq}(\bs{x},y)}{\partial y}+(x^{\eq}(\bs{x},y)-x^{\rm o})\right\}.
\end{align}
Because $\partial x^{\eq}(\bs{x},y)/\partial x_i\neq 0$, we obtain $\dot{\bs{x}}=\bs{0}\Leftrightarrow y=y^{\rm o}$. Further, when $y=y^{\rm o}$, we also obtain $\dot{y}=0\Leftrightarrow x^{\eq}(\bs{x},y)=x^{\rm o}$. \qed

\subsection*{Proof of Theorem 4}
We consider a neighborhood of the Nash equilibrium (i.e., $\bs{x}=\bs{x}^*+\rd\bs{x}$ and $y=y^*+\rd y$). Let $\bs{J}=(J_{ij})_{1\le i\le 5,1\le j\le 5}$ denote the Jacobian of learning dynamics;
\renewcommand{\arraystretch}{2.0}
\begin{align}
    J_{ij}:=\left\{\begin{array}{ll}
        \displaystyle \left.\frac{\partial\dot{x}_i}{\partial x_j}\right|_{\eqcon}=\left.x_i\tilde{x}_i\frac{\partial^2 u^{\eq}(\bs{x},y)}{\partial x_i\partial x_j}\right|_{\eqcon} & (1\le i\le 4,\ 1\le j\le 4) \\
        \displaystyle \left.\frac{\partial\dot{x}_i}{\partial y}\right|_{\eqcon}=\left.x_i\tilde{x}_i\frac{\partial^2 u^{\eq}(\bs{x},y)}{\partial x_i\partial y}\right|_{\eqcon} & (1\le i\le 4,\ j=5) \\
        \displaystyle \left.\frac{\partial\dot{y}}{\partial x_j}\right|_{\eqcon}=-\left.y\tilde{y}\frac{\partial^2 u^{\eq}(\bs{x},y)}{\partial y\partial x_j}\right|_{\eqcon} & (i=5,\ 1\le j\le 4) \\
        \displaystyle \left.\frac{\partial\dot{y}}{\partial y}\right|_{\eqcon}=-\left.y\tilde{y}\frac{\partial^2 u^{\eq}(\bs{x},y)}{\partial y^2}\right|_{\eqcon} & (i=j=5) \\
    \end{array}\right.\ .
\end{align}
\renewcommand{\arraystretch}{1.0}
From the definitions of $\dot{\bs{x}}$ and $\dot{y}$, we obtain
\begin{align}
    \left.\frac{\partial\dot{x}_i}{\partial x_j}\right|_{\eqcon}&=(u_1-u_2-u_3+u_4)\left.\left\{(y-y^{\rm o})\frac{\partial}{\partial x_j}\left(x_i\tilde{x_i}\frac{\partial x^{\eq}(\bs{x},y)}{\partial x_i}\right)\right\}\right|_{\eqcon}=0,\\
    \left.\frac{\partial\dot{x}_i}{\partial y}\right|_{\eqcon}&=x_i^*\tilde{x}_i^*(u_1-u_2-u_3+u_4)\left.\frac{\partial x^{\eq}(\bs{x},y)}{\partial x_i}\right|_{\eqcon}>0,\\
    \left.\frac{\partial\dot{y}}{\partial x_j}\right|_{\eqcon}&=-y^*\tilde{y}^*(u_1-u_2-u_3+u_4)\left.\frac{\partial x^{\eq}(\bs{x},y)}{\partial x_j}\right|_{\eqcon}<0,\\
    \left.\frac{\partial\dot{y}}{\partial y}\right|_{\eqcon}&=-2y^*\tilde{y}^*(u_1-u_2-u_3+u_4)\left.\frac{\partial x^{\eq}(\bs{x},y)}{\partial y}\right|_{\eqcon}<0.
\end{align}
We obtain the eigenvalues of matrix $\bs{J}$ as
\begin{align}
    &\mathrm{det}(\lambda\bs{E}-\bs{J})=0\\
    &\Leftrightarrow \lambda^3\left(\lambda^2-J_{55}\lambda-\sum_iJ_{i5}J_{5i}\right)=0\\
    &\Leftrightarrow \lambda=0\ ({\rm triple\ root}), \frac{J_{55}\pm\sqrt{J_{55}^2+4\sum_{i}J_{i5}J_{5i}}}{2}.
\end{align}
Here, $J_{55}< 0$ holds when $-\tilde{x}_1^*x_4^*+\tilde{x}_2^*x_3^*> 0$, while $J_{i5}J_{5i}<0$ always hold. Thus, each of the interior points of the Nash equilibrium has two negative eigenvalues, and its basin of attraction~[1] is two-dimensional in the five-dimensional space of $(\bs{x},y)$. Because the Nash equilibrium is spread over a three-dimensional space in $(\bs{x},y)$, the whole interior points of the Nash equilibrium are locally asymptotically stable (their basin of attraction is $3+2=5$-dimensional). \qed

\section{Experimental Results with More Samples} \label{AS02}
\begin{figure}[H]
    \centering
    \includegraphics[width=1.0\hsize]{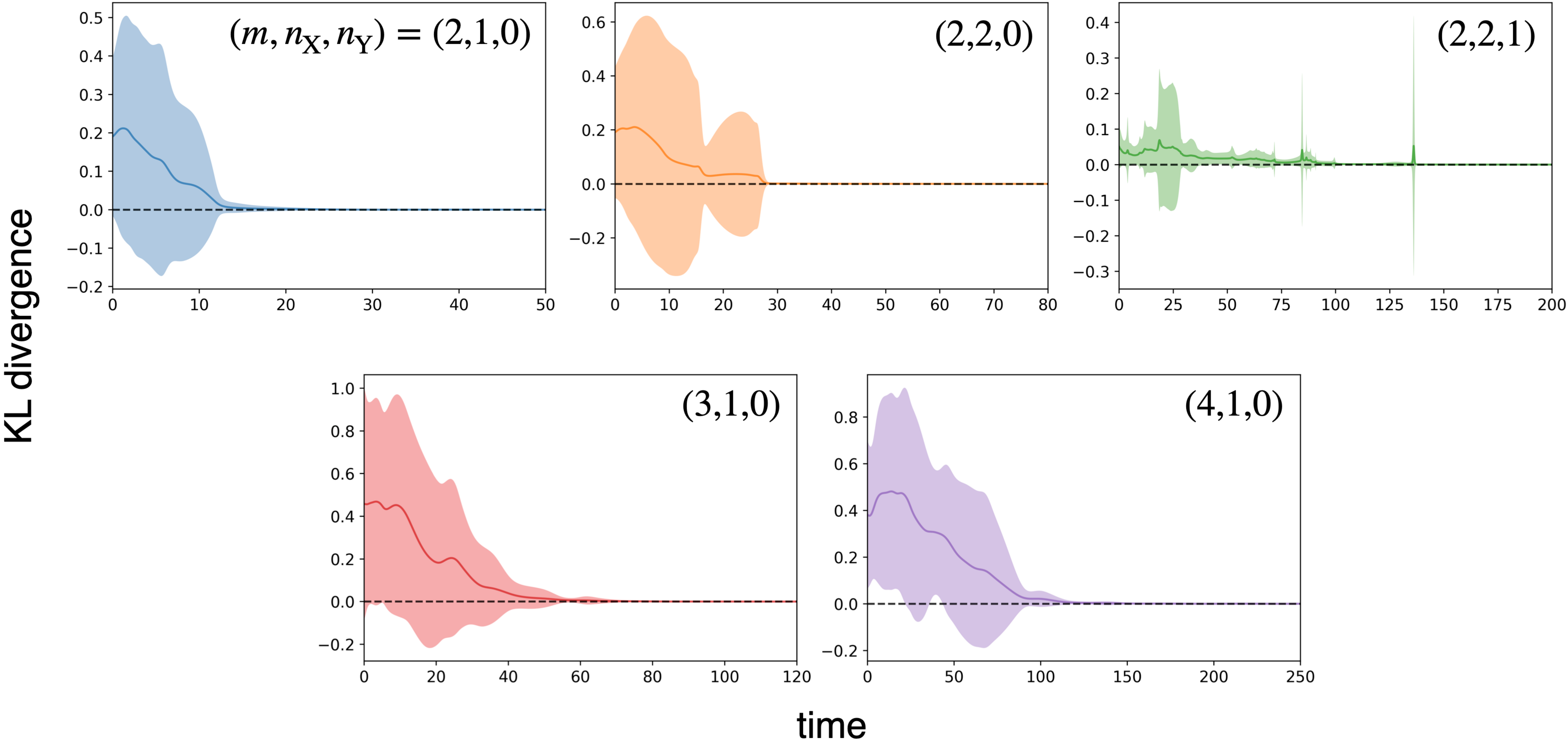}
    \caption{Experimental results with many samples. In each panel, the solid line shows the mean value of KL divergence for $50$ samples. The lightly colored area shows the standard deviation estimated from the $50$ samples. When $m=2$, we consider a matching-pennies game, where $x_a^{\rm o}=y_b^{\rm o}=1/2$ for all $a$ and $b$. When $m=3$, we consider a rock-paper-scissors game, where $x_a^{\rm o}=y_b^{\rm o}=1/3$ for all $a$ and $b$. When $m=4$, we consider an extended-rock-paper-scissors game, where $x_a^{\rm o}=y_b^{\rm o}=1/4$ for all $a$ and $b$.
    }
    \label{FS01}
\end{figure}

This section demonstrates that learning dynamics always converge to the Nash equilibrium, as shown in Fig.~A\ref{FS01}. We consider the five cases of $(m,n_{\rm X},n_{\rm Y})=(2,1,0)$ (blue panel), $(2,2,0)$ (orange), $(2,2,1)$ (green), $(3,1,0)$ (red), and $(4,1,0)$ (purple). 
We averaged $50$ samples of learning dynamics with random initial strategies for each case. 
As a measure for distance from equilibrium, we use the KL divergence to the marginalized strategies ${\bf z}^{\eq}:=({\bf x}^{\eq}, {\bf y}^{\eq})$ from the original Nash equilibrium ${\bf z}^{\rm o}:=({\bf x}^{\rm o}, {\bf y}^{\rm o})$;
\begin{align}
    D_{\rm KL}({\bf z}^{\rm o}\|{\bf z}^{\eq}):=\sum_{a\in\mc{A}}x_a^{\rm o}\log\frac{x_a^{\rm o}}{x_a^{\eq}}+\sum_{b\in\mc{B}}y_b^{\rm o}\log\frac{y_b^{\rm o}}{y_b^{\eq}}.
\end{align}
Here, the marginalized strategy is defined by
\begin{align}
    x_a^{\eq}:=\sum_{s\in\mc{S}}x_{a|s}p_{s}^{\eq}, \ \ \ y_b^{\eq}:=\sum_{s\in\mc{S}}y_{b|s_{n_{\rm Y}}}p_{s}^{\eq}.
\end{align}

From Fig.~A\ref{FS01}, let us discuss learning dynamics in detail. The more the number of memories (blue $<$ orange $<$ green) is and the more the number of actions (blue $<$ red $<$ purple) is, the longer time the convergence takes. This is because the number of variables that construct each agent's strategy increases with the number of memories and actions, and learning of the strategy gets slower with the increase in the number of such variables. Especially in the case of green panel $(m,n_{\rm X},n_{\rm Y})=(2,2,1)$, where both the agents have memories, complex dynamics are observed in a few samples; learning dynamics diverge from the Nash equilibrium once but converge there again.

% Figure S01

\section{Convergence Mechanism in $m$-Action Games} \label{AS03}
\begin{figure}[ht]
    \centering
    \includegraphics[width=0.7\hsize]{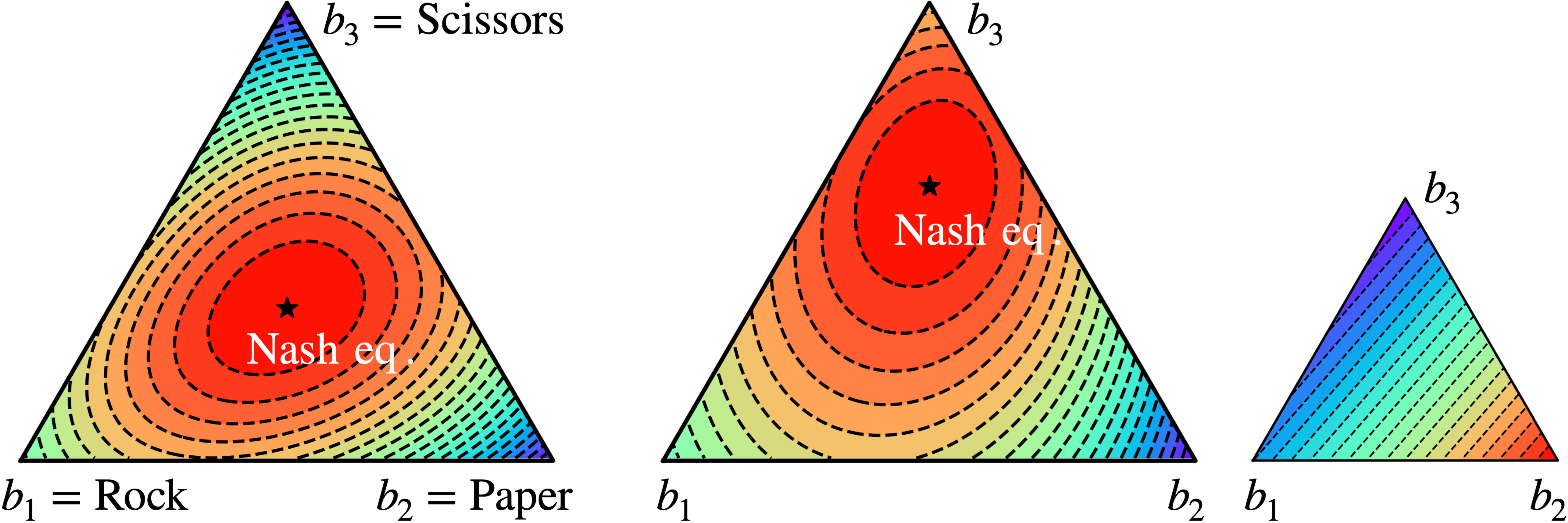}
    \caption{Under X's strategy fixed after it learns, Y's utility function is plotted depending on Y's strategy, i.e., ${\bf y}=\{y_1,y_2,y_3\}$. In all the panels, the red (blue) color indicates that Y's utility is large (small). The lower-left point of the simplex indicates that Y purely takes $b_1=$Rock action (i.e., $y_1=1$). The lower-right point indicates $b_2=$Paper action (i.e., $y_2=1$). The upper-middle point indicates $b_3=$Scissors action (i.e., $y_3=1$). In the ordinary (left) and weighted (center) rock-paper-scissors games, Y's utility function is strictly concave and takes its maximum value in the Nash equilibrium. In the right panel, X uses a $0$-memory strategy, and Y's utility function is linear.
    }
    \label{FS02}
\end{figure}

This section explains the convergence mechanism in $m$-action games between $1$-memory and $0$-memory players. In fact, Fig.~A\ref{FS02} shows that the longer-memory X learns to endow the other Y's utility function with a strict concavity in three-player zero-sum games, the ordinary (left panel) and biased (center) rock-paper-scissors games. Here, Y's utility function takes its maximum value at the Nash equilibrium, leading to convergence. Thus, the convergence to the Nash equilibrium is induced by the strict concavity of Y's utility function, as also seen in the case of two-action zero-sum games. If X has no memory, Y's utility function is always linear (right) and Y fails to converge to the Nash equilibrium.
% Figure S02

This strict concavity is given by a deadlock structure of actions, i.e., in rock-paper-scissors games, Rock $\to$ Paper $\to$ Scissors $\to$ Rock, and so on. Indeed, in Fig.~A\ref{FS02}, X tends to return Rock to the other's Scissors, Paper to Rock, and Scissors to Paper. No matter how biasedly Y chooses its action, its payoff is exploited by such a strategy of X. Thus, Y maximizes its utility at its minimax strategy and is induced to the Nash equilibrium.

By extending the numerical result given by Fig.~A\ref{FS02}, we establish a conjecture that the similar strictly concave utility function may be seen in general zero-sum games with full-support equilibrium. Such games exhibit a cyclical structure of advantage among actions, akin to Rock $\to$ Paper $\to$ Scissors $\to$ Rock, and so on. The strict concavity is maintained when a $1$-memory player consistently chooses advantageous actions in response to the other's previous choice of disadvantageous actions.

\section{Computational Environment} \label{AS04}
The simulations presented in this paper were conducted using the following computational environment.
\begin{itemize}
\item Operating System: macOS Monterey (version 12.4)
\item Programming Language: Python 3.11.3
\item Processor: Apple M1 Pro (10 cores)
\item Memory: 32 GB
\end{itemize}

\section*{References}
% wiggins2003introduction
[1] Stephen Wiggins. {\it Introduction to applied nonlinear dynamical systems and chaos}, volume 2. Springer, 2003

\end{document}